\documentclass[pdftex,twocolumn,epjc3]{svjour3}          

\usepackage{caption}
\usepackage{subcaption}
\usepackage{amsmath}

\RequirePackage[T1]{fontenc}

\smartqed  

\RequirePackage{graphicx}
\RequirePackage{mathptmx}      
\RequirePackage{flushend}
\RequirePackage[numbers,sort&compress]{natbib}
\RequirePackage[colorlinks,citecolor=blue,urlcolor=blue,linkcolor=blue]{hyperref}

\journalname{Eur. Phys. J. B}

\begin{document}

\title{Searching for clusters of  targets under stochastic resetting}


\author{Georgia R. Calvert \thanksref{addr1}        \and
        Martin R. Evans \thanksref{e1,addr1} 
}

\thankstext{e1}{e-mail: m.evans@ed.ac.uk}

\institute{
SUPA, School of Physics and Astronomy, University of Edinburgh, Peter Guthrie Tait Road, Edinburgh EH9 3FD, United Kingdom
\label{addr1}
}

\date{Received: date / Accepted: date}

\abstractdc{
We consider diffusion under stochastic resetting to the origin in one dimension and compute the mean time
to find both of two targets placed either side of the origin. A surprising result is that  increasing the distance between two targets can \emph{decrease} the overall search time. We compute the optimal arrangement of two targets in limiting cases.
 We generalise to obtain recursive expressions for  the mean time to find all of  multiple targets. We discuss the relevance to real-world
problems of locating multiple targets such as  proteins locating clusters of  DNA lesions.
}

\maketitle

\section{Introduction}

It has long been appreciated that search processes in biology are speeded up by the use search strategies which include  long-range  as well as short-range moves \cite{VdLRS11}.
For example, in order that vital biological functions occur, proteins must locate binding sites on DNA rapidly in order to trigger various transcription processes. In fact they locate sites up to 1000 times faster than expected for a diffusion controlled process \cite{ref:ber}. The first theories that attempted to account for these fast search times proposed {\em facilitated diffusion} as the search mechanism \cite{ref:hip1,ref:hip2,ref:hip3}. This mechanism reduces the dimensionality of the search by splitting it into 1D and 3D components: the proteins search in 1D along a strand of DNA to which they are loosely bound and then disassociate, diffuse in 3d and re-associate at another non-specific site, which may be far along the strand, to continue the 1D search. 

More recently,   search processes have  been of interest within the statistical physics  community 
and different classes of search strategies have been identified, see e.g. \cite{MZ02,BCMSV05,Gelenbe10,Snider12}.
Different  specific search problems may have  different protocols, but they share the desire for an optimal search strategy.  In an intermittent search (see \cite{BLMV11} for a review)
there are local steps, which effect searching,  interspersed with long distance  relocations. Such a combination of moves is believed to be beneficial  in  wide range of biological behaviours from  animal foraging to  target search of proteins on DNA molecules \cite{CBVM04,LKMK08,BKSV09}.

A simple model for searching with short-range and long-range moves is diffusion with stochastic resetting \cite{ref:eva}. Here short range foraging is modelled by diffusion of the searcher  and long-range relocations by instantaneous `reset events'. In the simplest framework the resetting of the searcher is to a fixed resetting location but a distribution of resetting sites may also be considered \cite{ref:eva2}. It has been shown that the mean time for the searcher  to locate  a fixed target  is significantly improved by the inclusion of resetting. Indeed, for a purely diffusive process the mean time to locate a target diverges whereas it is rendered finite by the introduction of a resetting process. Moreover, by tuning the resetting rate, one may optimise the mean time to locate a target.
The resetting paradigm has been explored in a number of other contexts including restarting of stochastic algorithms\cite{Lorenz18} and complex chemical reactions\cite{RUK14}---see \cite{ref:eva3} for a recent review.
Recently, experiments in optical traps have  measured
the optimal mean time for a diffusing Brownian particle to reach a target under resetting \cite{BBPMC20, TSSR20}.

Diffusion with stochastic resetting  has  the appealing feature that many properties may be analysed exactly.
Mostly, this has been carried out in one spatial dimension \cite{ref:eva} but one can easily extend some results to higher spatial dimensions \cite{EM14}. Typically, one focusses on the mean time for a searcher  to find a target at which point the searcher is absorbed. 
Some works have considered multiple targets \cite{CS18,PP19,Bressloff20c,Bressloff20a, Bressloff20b}.
However, in some applications it may happen that there are a number of targets, distributed in some manner, and the goal is to locate {\em all} of the targets. A  real world context  is that of proteins searching for DNA lesions \cite{georgia}. DNA lesions need to be located by proteins quickly in order to be fixed, If there are many DNA lesions and proteins are successfully finding just one of them, then the bulk of the tissue will remain damaged so a  successful search must try to locate them all.
Moreover, if the tissue has been exposed to a large dose of ionising radiation which creates the DNA lesions, the  lesions may exist in clusters \cite{ref:sun,ref:sag,ref:nic}.
In order to address the problem of locating each and every one of a cluster of targets, the logical first step is to consider a searcher looking for two targets, rather than just one.

In this paper, inspired by the  problem of locating all of a cluster of targets, we consider the problem of diffusion with resetting with a single searcher and the mean time to locate multiple targets in one dimension.
We find exact expressions for  the mean time to multiple abosorption. An interesting emergent effect  is that the presence of multiple targets combined with resetting after locating  each of them, can lead to a {\em reduction} in the mean time to find a single target. This counterintuitve result illustrates that co-operative effects may arise from
having a cluster of targets to find.

The paper is organised as follows. In section 2 we  define the model beginning with two targets on either side of a resetting site and set up the formalism for the computation of mean time to  double absorption. In section 3 we present results for the Laplace transform of the survival probability and the mean time to double absorption. In section 4 we generalise these results to an arbitrary number of targets and present a recursive formula for the mean time to multiple absorptions.
In section 5 we conclude with some comments on how the model may be improved with respect to the real-world problem of  modelling proteins locating DNA lesions.

\section{Model definition: One Searcher, Two Targets}
We consider a diffusive particle (searcher) with diffusion constant $D$ moving on a one-dimensional lattice. With rate $r$ the searcher is reset instantaneously to the origin.
We consider one target at $x_l<0$ and the other at $x_r>0$. 
When  the searcher touches a target for the first time, the target is absorbed and the searcher is   instantaneously restarted at the origin.
It is important to note that once either $x_r$ or $x_l$ is located, it is no longer an absorbing target i.e. if the searcher touches the target again, the searcher is not restarted.
The search is completed when {\em both} targets have been located.

We note that a related problem of diffusion with resetting on a one-dimensional domain with absorbing boundaries has been considered in \cite{PP19,PP19b} where the statistics of the  time to be absorbed by {\em either} boundary were considered.

Our goal is to find the mean time to find both targets (mean time to double absorption, MTDA).
For usual first passage problems there are a variety of approaches to calculating survival probabilities and mean first-passage times: forward Fokker-Planck equation, backward Fokker Planck equation and renewal equations (see e.g. \cite{ref:red,BMS13,ref:eva3}).  Here we find it most convenient to use the first approach.

There are three relevant survival probability densities to consider:

\noindent $q(x,t)$:  the probability density of finding the searcher at position $x$, time $t$ with it not having touched either $x_r$ or $x_l$.

\noindent $q_r(x,t)$: the probability density for the searcher to be at position $x$ at time $t$ and it having touched  the target at $x_r$ but not $x_l$.

\noindent $q_l(x,t)$: the probability density for the  searcher to be at position $x$ at  time $t$ and it having touched the  target at  $x_l$  but not $x_r$.

The  survival probability density, $q(x,t)$, satisfies the forward master equation,  in which the spatial variable $x$ is the position after time $t$:
\begin{multline}\label{orig}
    \frac{\partial{q(x,t)}}{\partial{t}} = -rq(x,t) + D\frac{\partial^2q(x,t)}{\partial{x^2}} \\
 + r\left[ \int^{x_r}_{x_l} dx \; q(x,t) \right] \delta(x)\;,
\end{multline}
with boundary conditions,
\begin{equation}
    q(x_l,t)=q(x_r,t)=0\;.
\label{qbc}
\end{equation}
The  initial condition is
\begin{equation}
    q(x,0)=\delta(x)\, ,
\end{equation}
since the searcher begins at the origin at $t=0$.

The first term on the right hand side of  (\ref{orig})  represents a loss of probability at $x$ due to resetting and the third term on the right hand side indicates a gain of probability at the origin due to resetting from all $x$ within the domain. The boundary conditions (\ref{qbc}) correspond to absorption of the searcher when it touches $x_r$ or $x_l$.
One can write down analogous but more involved equations for $q_r(x,t)$ and $q_l(x,t)$, but as we shall see we do not need to explicitly compute these quantities. 

The rate at which the search is completed when the second of the targets   is found (with one target already having been found)  is the  sum of the rate of locating $x_l$ once $x_r$ already been found and the rate of locating $x_r$ once $x_l$ has already been found.
These two rates are given by the diffusive currents from $q_r$ at $x_l$
and from $q_l$ at $x_r$, respectively.
 Thus, the rate at which the second of the targets is located, and the search completed, can   be written as 
\begin{equation}
    F(t) = D \frac{\partial{q_r(x,t)}}{\partial{x}}\Big|_{x=x_l} - D \frac{\partial{q_l(x,t)}}{\partial{x}}\Big|_{x=x_r}\;.
\end{equation}

Now, the rate of locating the final target may also be written as the negative rate of change of the total survival probability, $Q_{\rm tot}(t)$,
\begin{equation}
    F(t) = 
-\frac{\partial Q_{\rm tot}(t)}{\partial{t}} 
\label{rate2}
\end{equation}
where
\begin{equation}
Q_{\rm tot}(t) =\int^{\infty}_{x_l}dx \, q_r(x,t) + \int^{x_r}_{-\infty}dx \, q_l(x,t) + \int^{x_r}_{x_l} dx \, q(x,t).\label{Qdef}
\end{equation}
is the total probability that the searcher has not yet touched {\em both} targets.
Note that this survival probability contains three terms:  the probability of having touched $x_r$ but not $x_l$; the probability of having touched $x_l$ but not $x_r$;
and the probability of having touched neither target.
The limits on the integrals on the rhs of (\ref{Qdef})  are distinct because for each survival probability density the allowed $x$ domain is different e.g. for $q_r(x,t)$, $x$ can vary from $x_l$ to $\infty$ since once $x_r$ has been touched it is no longer absorbing.

In order to obtain the MTDA, the standard approach  for mean first passage time calculations
would be to average the time to double absorption over the rate of absorption, $F(t)$, to obtain
\begin{equation}
    T_2 = \int^{\infty}_{0} dt \; tF(t) =\int^{\infty}_{0} dt \, Q_{\rm tot}(t)\, ,
\label{TQ}
\end{equation}
after integrating by parts, assuming the survival probabilities decay faster than $1/t$. We use the notation $T_2$ to emphasize that 
it is the mean time to find both targets.
Defining  the Laplace transforms of the survival probabilities as
\begin{eqnarray}\label{LT}
    \widetilde{q}(x,s) &=& \int^{\infty}_{0}dt \; e^{-st}q(x,t)\\
  \widetilde{q}_r(x,s) &=& \int^{\infty}_{0}dt \; e^{-st}q_r(x,t)\\
  \widetilde{q_l}(x,s) &=& \int^{\infty}_{0}dt \; e^{-st}q_l(x,t)\, ,
\end{eqnarray}
 equation (\ref{TQ}) can be written
\begin{equation}
    T_2 = \int^{\infty}_{x_l}dx \; \widetilde{q_r}(x,0) + \int^{x_r}_{-\infty}dx \; \widetilde{q_l}(x,0) + \int^{x_r}_{x_l} dx \; \widetilde{q}(x,0)\,.
\end{equation}
The standard approach would  therefore be to compute the Laplace transforms (\ref{LT}) , through which   the MTDA will ultimately be found.

To shorten the calcuation, we will take advantage of the known result  for $T_1(X_r)$, the mean time to absorption for diffusion under resetting to the origin with a single target at $X_r$ \cite{ref:eva},
\begin{equation}
T_1(X_r)=
\frac{1}{r}\left({\rm e}^{\alpha |X_r| }-1 \right)\,.
\label{T1}
\end{equation}

For our case  the MTDA, $T_2(x_l,x_r)$, can be written in terms of the probabilities $P_r$, $P_l$ to find the right, left target first, the mean times $T_r$, $T_r$ to find the corresponding target, conditioned on that target being found first, and the mean time to find a single target, once the other  target
 has been eliminated
\begin{eqnarray}
T_2(x_r,x_l) &=& P_r\left[ T_r + T_1(x_l)\right] +  P_l\left[ T_l + T_1(x_r)\right] \\
&=& \tau + P_rT_1(x_l)+  P_l  T_1(x_r) \label{Tfast}
\end{eqnarray}
where we have defined
\begin{equation}
\tau = P_rT_r +   P_lT_l ,  \label{tau}
\end{equation}
which is the mean first passage time to {\em either} of the targets at $x_r$, $x_l$.
 We note that $P_r$, $P_l$ are referred to in the literature as splitting probabilities
and have been studied in  \cite{Belan18,PP19}. The relation \eqref{tau} was used in \cite{PP19}.
Thus  (\ref{Tfast}) reads that the mean time to double absorption is the mean time to the first absorption plus the
average of the mean time for the second absorption weighted according to the probabilities of which absorption occurs first.
The quantities $\tau$, $P_r$, $P_l$ appearing in (\ref{Tfast}) only require the knowledge of 
$\widetilde{q}(x,0)$.
To see this note that
\begin{equation}
   \tau  = \int^{\infty}_{0} dt \, \int_{x_l}^{x_r} dx\, q(x,t) = \int_{x_l}^{x_r} dx\, \widetilde{q}(x,0)
\label{tauint}
\end{equation}
and $P_l$ is given by the integral of the absorption rate at target $x_l$
\begin{equation}
P_l = \int_0^\infty dt\, D \frac{\partial q(x,t)}{\partial x} \Big|_{x_l} = D \frac{\partial \widetilde{q}(x,0)}{\partial x} \Big|_{x_l}\; .
\end{equation}
Similarly,
\begin{equation}
P_r = -\int_0^\infty dt\, D \frac{\partial q(x,t)}{\partial x} \Big|_{x_r} = -D \frac{\partial \widetilde{q}(x,0)}{\partial x} \Big|_{x_r}\; .
\end{equation}

Our task is therefore to compute $\widetilde{q}(x,0)$.  For completeness we compute in the appendix the full Laplace transform  $\widetilde{q}(x,s)$.

\section{Exact expressions for Laplace transform of survival probability and MTDA}

\subsection{ Laplace transform of survival probability $\widetilde{q}(x,s)$}
The Laplace transform of equation (\ref{orig})  is
\begin{multline}
\left[q(x,t)e^{-st}\right]^{\infty}_{0}+s\widetilde{q}(x,s) = -r\widetilde{q}(x,s)+D\frac{\partial^2\widetilde{q}(x,s)}{\partial{x^2}}\\
+r\left[\int^{x_r}_{x_l}dx \; \widetilde{q}(x,s)\right]\delta(x)\;.
\end{multline}
Rearranging yields
\begin{equation}\label{lap}
-(r+s)\widetilde{q}(x,s)+D\frac{\partial^2\widetilde{q}(x,s)}{\partial{x^2}}  = -E\delta(x)\,,
\end{equation}
where
\begin{equation}
E = 1+r\left[\int^{x_r}_{x_l}dx \; \widetilde{q}(x,s))\right]\,.
\label{Edef}
\end{equation}
In the appendix we give details of the solution of (\ref{lap},\ref{Edef}) for $\widetilde{q}(x,s)$.
Here we present the result for $\widetilde{q}(x,0)$, which we use in the following,
\begin{eqnarray}\label{di1}
    \widetilde{q}(x,0)&=&\frac{\sinh \alpha_0 x_r \sinh \alpha_0 (x-x_l)}{\alpha_0{D}(\sinh{\alpha_0{x_r}}-\sinh{\alpha_0{x_l}})} \quad\mbox{for} \;\; x<0 \\
\label{di2}
\widetilde{q}(x,0)&=& \frac{\sinh \alpha_0 x_l \sinh \alpha_0 (x-x_r)}{\alpha_0{D}(\sinh{\alpha_0{x_r}}-\sinh{\alpha_0{x_l}})}\quad \mbox{for}\;\; x>0,
\end{eqnarray}
where
\begin{equation}
\alpha_0 = \left(\frac{r}{D}\right)^{1/2}
 \end{equation}

\subsection{Mean Time to Double Absorption, $T_2$}

We now put everything together to obtain an expression for the mean time to double absorption (MTDA) using (\ref{Tfast}).
First we compute $\tau$, the mean time to absorb the first target, from the formula (\ref{tauint}) using expressions
(\ref{di1},\ref{di2}) to  obtain
\begin{equation}
\tau = \frac{1}{r} \left[ \frac{\sinh \alpha_0(x_r-x_l)}{\sinh\alpha_0 x_r - \sinh \alpha_0 x_l} -1 \right]\,.
\end{equation}
This expression is in agreement with equation (15) of \cite{PP19}.
One can check that the limits $x_{r} \to\infty$ and $x_l \to -\infty$ recover (\ref{T1}).

We also require the following  expressions
\begin{eqnarray}
    \frac{\partial\widetilde{q}(x,0)}{\partial{x}}\Big|_{x=x_r} &=&\frac{\sinh \alpha_0 x_l}{D(\sinh{\alpha_0{x_r}}-\sinh{\alpha_0{x_l}})}\\
    \frac{\partial\widetilde{q}(x,0)}{\partial{x}}\Big|_{x=x_l}&=&\frac{\sinh \alpha_0 x_r}{D(\sinh{\alpha_0{x_r}}-\sinh{\alpha_0{x_l}})}\;,
\end{eqnarray}
which yield
\begin{eqnarray}
    P_r &=& \frac{-\sinh \alpha_0 x_l}{(\sinh{\alpha_0{x_r}}-\sinh{\alpha_0{x_l}})}\\
    P_l&=&\frac{\sinh \alpha_0 x_r}{(\sinh{\alpha_0{x_r}}-\sinh{\alpha_0{x_l}})}\;.
\end{eqnarray}
These expressions are in agreement with equations (31,32) of \cite{PP19}.

We then obtain from  using (\ref{Tfast}), after simplification,
\begin{multline}\label{yes}
    T_2 = \\
\frac{1}{r}\left(\frac{\sinh^2{\alpha_0{x_r}}+\sinh^2{\alpha_0{x_l}}}{\sinh{\alpha_0{x_r}}-\sinh \alpha_0{x_l}}+\cosh{\alpha_0{x_r}}+\cosh{\alpha_0{x_l}}-2\right)
\end{multline}
Expression (\ref{yes}) is the main result of this section.

One can check that as $x_r \to 0$ one obtains
\begin{equation}
T_2 \to 
\frac{1}{r}\left({\rm e}^{-\alpha_0 x_l }-1 \right)
\end{equation}
which recovers the  mean time to absorption under stochastic resetting  for a single target at $x_l$ (\ref{T1}). The reason is that as $x_r \to 0$ (or $x_l \to 0$)  one target is immediatly located and eliminated by the searcher starting from the origin. Then the searcher effectively starts again
 from origin at $t=0$ to search for a single target.

We have also checked (\ref{yes}) by the longer route of computing $\widetilde{q}_r(x,0)$, $\widetilde{q}_l(x,0)$
and using formulas (\ref{TQ}), (\ref{Qdef})  and found  perfect agreement.

\subsection{Dependence of  MTDA on $r$}

\begin{figure}[h]
\centering
\begin{subfigure}{1.0\columnwidth}
  \centering
  \includegraphics[width=0.75\columnwidth, angle =270]{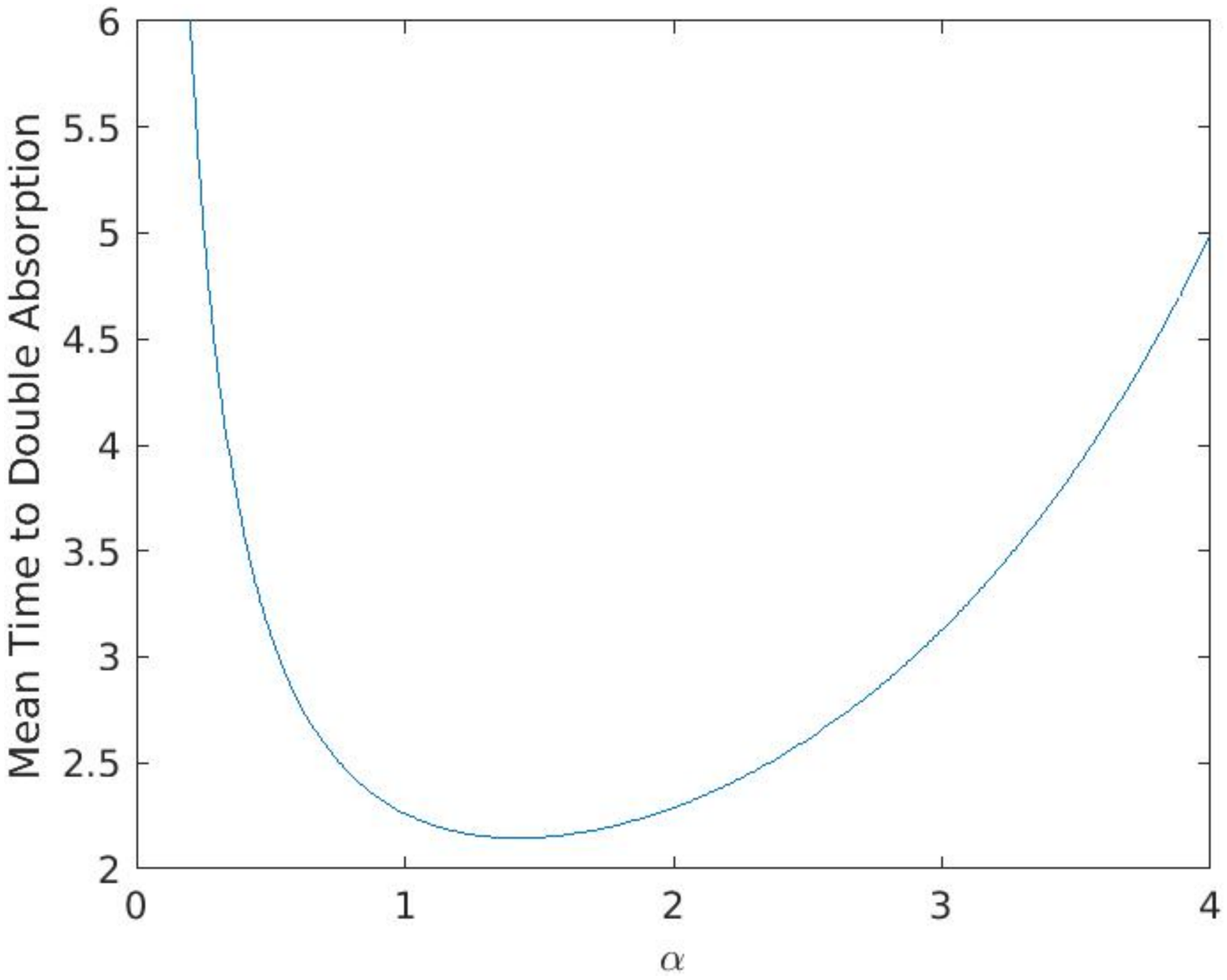}
  \caption{Mean time to double absorption plotted against $\alpha_0$, defined in equation (\ref{alp}).}
  \label{fig:sub1}
\end{subfigure}%
\qquad
\begin{subfigure}{1.0\columnwidth}
  \centering
  \includegraphics[width=0.75\columnwidth, angle=270]{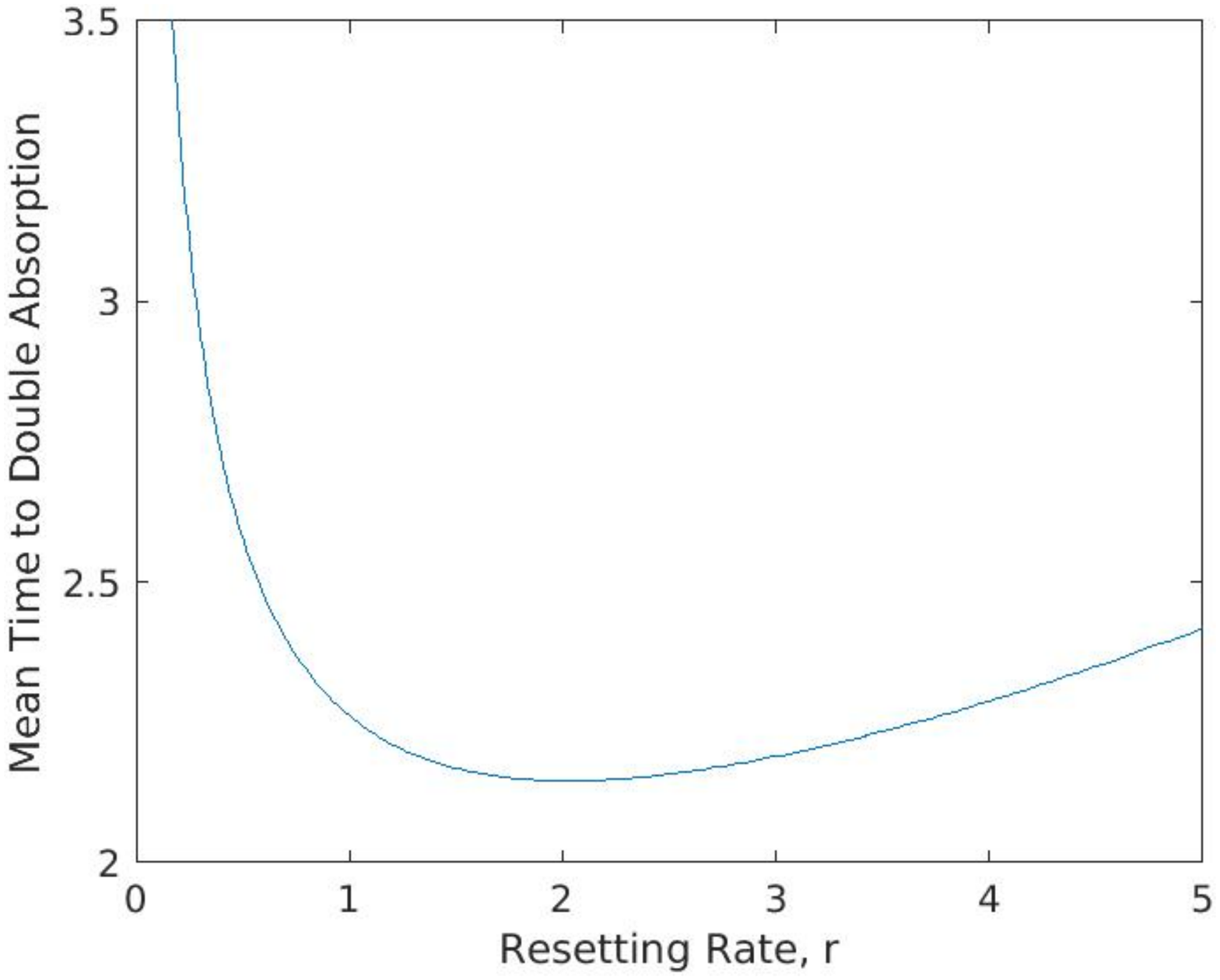}
  \caption{Mean time to double absorption plotted against resetting rate, $r$.}
  \label{fig:sub2}
\end{subfigure}
\caption{Plots of mean time to double absorption with $D=1$, $x_l=-1$ and $x_r=1$. There is an obvious point at which MTDA is minimised. It is at this value of $r$ that the resetting rate is optimised.}
\label{fig:a_r}
\end{figure}

Plots of equation (\ref{yes}) with $T$ plotted against both $\alpha_0$ and $r$ (with  parameter values $D=1$, $x_l=-1$ and $x_r=1$) exhibit the expected features (Figure \ref{fig:a_r}): \\
(i) MTDA tends to infinity as the resetting increases as the searcher does not have sufficient time to diffuse far enough to find a target between resets and therefore will never find either target.\\
(ii) MTDA tends to infinity as the resetting rate tends to zero. This is because, after having found the first target (inevitable given   the searcher is sandwiched between the two targets, and there is no resetting),
the system reduces to  the well-studied system of one diffusive particle searching in 1D for one fixed target without resetting \cite{ref:eva}. In this system, the mean time to absorption diverges.\\
(iii) There is a turning point at which MTDA is minimised (see Fig. 1 \ref{fig:sub2}). The value for $r$ at this minimum is the optimal value for $r$; for this set of parameters optimal $r$ is $\simeq 2.03$ for which MTDA is $\simeq 2.15$.

One can ask what is the optimal resetting rate i.e. the value of $r$ that minimises (\ref{yes}). For the case of a single target at $X_r$ it was found that there is a unique minimum value, which is most neatly expressed in terms of the variable $\gamma = X_r \alpha_0$ which is the ratio of distance to target over the typical length diffused between resets. For the two-target case considered here, the problem is more complicated since there are two lengths $x_r$, $|x_l|$.
The simplest case to consider is $ |x_l| = x_r$ where again we can express the minimisation in terms of a single variable 
\begin{equation}
\gamma = x_r \alpha_0\;.
\end{equation}
Equating the derivative of (\ref{yes}) with respect to $r$ to zero, yields the following transcendental equation
\begin{equation}
3(\gamma-2)\exp(\gamma) - (\gamma +2) \exp(-\gamma) + 8 =0\;.
\end{equation}
The unique non-zero solution is $\gamma  = 1.42433$, to be compared to $\gamma = 1.59362$ for the case of a single target at $x_r$ \cite{ref:eva}. For the case $D=1$, $x_r=1$ this equates to an optimal  resetting rate $r= 2.02872$, which is in agreement with Fig. 1b). The fact that the optimal  value of $\gamma$ (and consequently of $r$) is lower for double absorption than for a single target is easy to understand.
In the case of two equidistant targets it would be optimal to have zero resetting rate up until the first target is found and then adopt the optimal resetting rate for a single target search. This suggests that  a lower constant  resetting rate is optimal overall.

\subsection{Dependence of  MTDA on $x_r$}
We now turn to the dependence of the MTDA on the positions of the targets
\begin{figure}[h]
\centering
\includegraphics[width=0.75\columnwidth, angle=270]{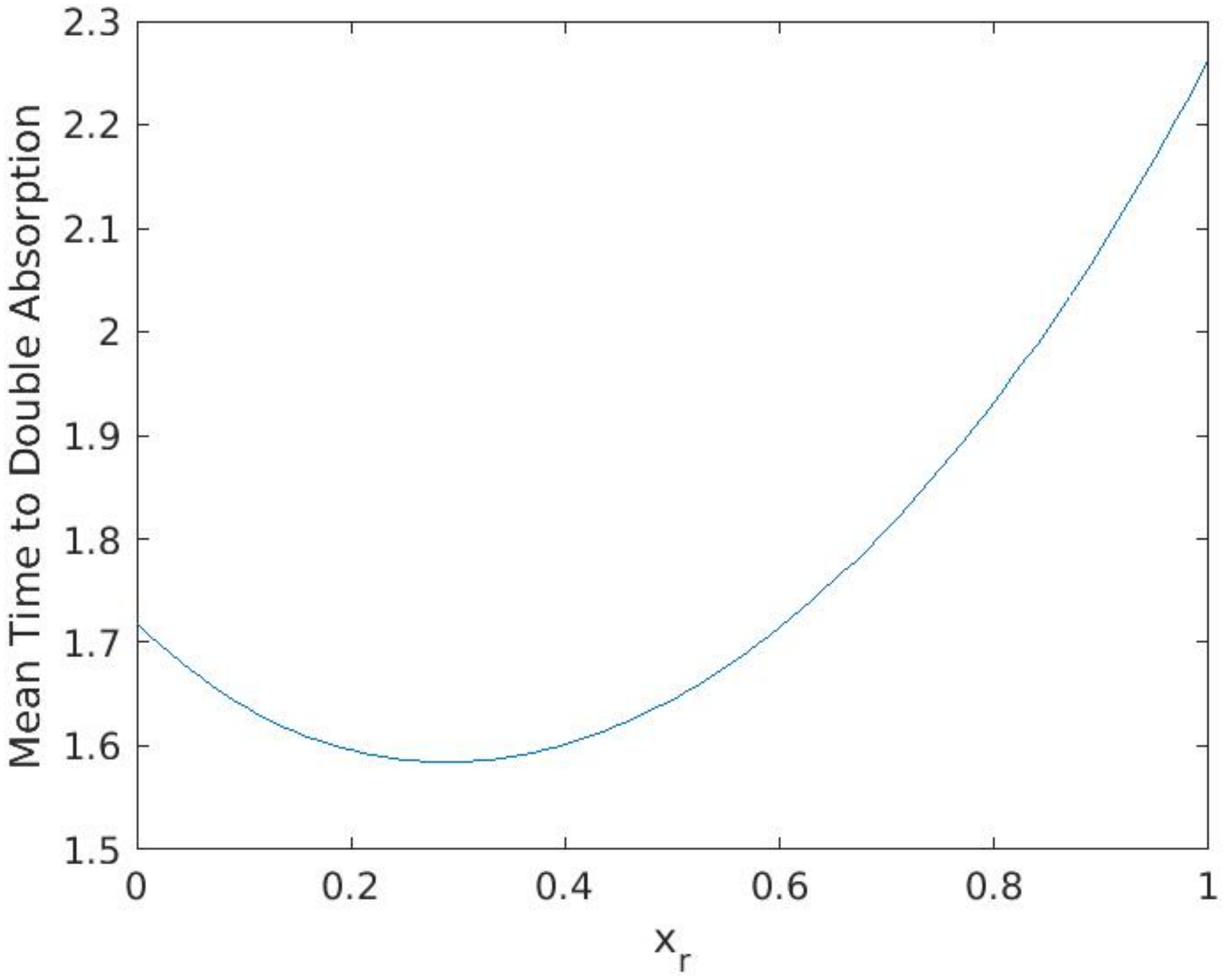}
\caption{Mean time to double absorption plotted against $x_r$, with $D=1$, $r=1$ and $x_l=-1$. The minimum occurs at $x_r =0.289$.}
\label{fig:x_r}
\end{figure}



Plotting $T$ against $x_r$ (again with arbitrary parameters $D=1$, $x_l=-1$ and $r=1$) produces a surprising result (Figure \ref{fig:x_r}). Whilst the plot exhibits the expected feature of a divergent MTDA as $x_r$ tends to infinity, the position of the turning point is unexpected. With the target at $x_r$ able to exist at any point $x_r>0$ and the target at $x_l$ fixed at $x_l=-1$, naively one might assume that the MTDA would be minimised when $x_r=0$. In this scenario, $x_r$ would be found immediately, instantly resetting the system followed by a search for $x_l$ (essentially, this scenario is \emph{only} a search for one target, $x_l$). However, placing $x_r$ \emph{further} from $x_l=-1$ actually \emph{reduces} the MTDA. There exists a kind of {\em cooperative effect} in which the existence of a second target makes the search for the first target quicker.

The reason for this lies in the fact that the searcher is reset to the origin once it finds a target. Therefore having a target at $x_r>0$ actually cuts off some trajectories that are moving away from $x_l$.
However, placing $x_r$ too far from the origin negates the cooperative effect. If $x_r$ becomes too large, the time taken to find $x_r$ increases such that it cancels out the benefit from the cooperative effect. Thus there exists an optimum value for $x_r$.

To check that cooperative effect between targets  always occurs one can compute
\begin{equation}
\frac{\partial T_2}{\partial x_r} \Big|_{x_r =0} = -\frac{\alpha_0}{r}\;.
\end{equation}
Thus, there is a decrease in $T_2$, on increasing $x_r$ from zero, for all parameter values.

 We now  consider the value of $x_r>0$ that minimises the MTDA given a fixed  $x_l <0$.  Setting the derivative of \eqref{yes} with respect to $x_r$ to zero yields
a quartic equation for $\eta \equiv {\rm e}^{\alpha_0 x_r}$:
\begin{equation}
(\eta^2 - 1)^2 + 4 \sinh(\alpha_0 |x_l|)\eta(\eta^2-1) -4 \sinh^2(\alpha_0 |x_l|)=0\;.
\end{equation}
The form of this  equation implies that for $\eta >1$  (which corresponds to $x_r>0$) there is  always a unique solution. Thus, there is a unique optimal value  $x^*_r$ of $x_r$.

We can obtain the behaviour of $x^*_r$ in the  two limiting cases where $\alpha_0 |x_l|$ is either large or small. The quantity  $\alpha_0 |x_l|$ is the ratio of the distance to the left target from the resetting site to the typical distance diffused between resets \cite{ref:eva}.\\[1ex]

\noindent{\bf For $\alpha_0 |x_l|  \gg 1$}
\begin{equation}
 {\rm e}^{\alpha_0 x^*_r} \simeq \left( \frac{ {\rm e}^{\alpha_0 |x_l|}}{2}\right)^{1/3},
\end{equation}
so 
\begin{equation}
x^*_r \simeq \frac{|x_l|}{3}\;.
\end{equation}
Also the probability of locating the right target first is
\begin{equation}
P_r \simeq 1 - \frac{1}{2^{1/3}}{\rm e}^{-2 \alpha_0 |x_l|/3}
\end{equation}

\noindent{\bf For $\alpha_0 |x_l|  \ll 1$}
\begin{equation}
 \left({\rm e}^{\alpha_0 x^*_r}-1 \right) \simeq \left( 2^{1/2}-1\right) \sinh \alpha_0 |x_l |\;,
\end{equation}
so 
\begin{equation}
x^*_r \simeq \left( 2^{1/2}-1\right) |x_l|\;.
\end{equation}
Also the probability of locating the right target first is
\begin{equation}
P_r \simeq 2^{-1/2} =  0.7071\ldots
\end{equation}

It is interesting to note that in both limiting cases the optimal distance of the right target is  less than the distance to the left target, by simple factors of $1/3$ for $\alpha_0 |x_l|  \gg 1$ and  $0.4142\ldots$ for
$\alpha_0 |x_l|  \ll 1$. Also, the probability of locating the right target first is greater than one half in both cases. Thus in its optimal position the right target effectively cuts off errant trajectories.

\section{Generalisation to multiple targets}

Here we outline how the results may be generalised to an arbitrary number of targets on the  real line.
So far we have considered two targets either side of the resetiing position, the origin.
The result for two targets on the same side, say at positions $x_{r_1}>0$ and $x_{r_2} >x_{r_1}$ is simply
\begin{equation}
T_2 (x_{r_1}, x_{r_2}) = T_1(x_{r_1}) + T_1(x_{r_2})\,,
\end{equation}
i.e. it is the mean time to locate the nearest target plus the mean time to find the furthest target.
For the case of two targets to the right of the origin,  $x_{r_1}>0$ and $x_{r_2} >x_{r_1}$ and one to the left $x_l <0$, using the same logic as for (\ref{Tfast}), the mean time to find all three targets is
\begin{equation}
T_3 (x_l, x_{r_1}, x_{r_2}) = \tau(x_l, x_{r_1})+P_{r_1} T_2(x_l,x_{r_2})  + P_lT_2(x_{r_1},x_{r_2}).
\label{T3}
\end{equation}
Equation (\ref{T3}) states   that the mean time to triple  absorption is equal to  the mean time to the first absorption plus the
average of the mean time for the remaining double absorption weighted according to the probabilities of which absorption occurs first.

In this way, one can recursively write down expressions for the mean time to absorb any number of targets. Specifically, for $m$ targets to the left of the origin and $n$ targets to the right the expression reads
\begin{multline}
T_{m+n} (x_{l_m}, \dots x_{l_1}, x_{r_1}, \dots ,x_{r_n}) = \tau(x_l, x_{r_1})\\
+P_{r_1}(x_{l_1},x_{r_1}) T_{n+m-1}(x_{l_m}, \dots x_{l_1}, x_{r_2}, \dots ,x_{r_n}) \\  
+P_{l_1}(x_{l_1},x_{r_1}) T_{n+m-1}(x_{l_m}, \dots x_{l_2}, x_{r_1}, \dots ,x_{r_n}) \;, \label{Tmulti}
\end{multline}
where $P_{r_1}(x_{l_1},x_{r_1})$  is now  the probability of first finding the target on the right of the pair of closest targets on either side of the origin.
\section{Conclusion}

In this paper we have studied mean times for a diffusive searcher, under resetting to a fixed position, to locate all of  multiple targets.
We have presented the exact expression (\ref{yes}) for two targets in one dimension and shown how to generalise to an arbitrary number of targets on the real line (\ref{Tmulti}).

A perhaps surprising result is that  the presence of multiple targets can actually reduce the time to find a single target. In the case of two targets we have obtained expressions for the optimal position
of the second target which minimises the time to locate both targets. Although seemingly counter-intuitive,  the effect is a result of the searcher being returned to the origin once a target has been located.

The  study  was motivated by the problem of healing of lesions on DNA. 
For this purpose the model we consider is  necessarily a crude simplification
where each event of disassocation/re-association from the DNA is considered a reset of the system and the resetting instantaneously occurs at a single resetting site.
In reality there are many possible processes for the translocation of proteins between sites e.g. {jumping}, {hopping}, {intersegment transfer} and {sliding}
are commonly discussed \cite{ref:hip3}. With the emergence of single molecule methods, it has become possible to observe the motion of proteins at an individual molecule level \cite{ref:gor}, which may inform modelling.
An obvious  improvement to be made to the resetting dynamics is  to more faithfully model a  distribution of binding sites (for resetting to) and to include a delay in the reset process. A cluster of target sites may also have some specific structure.
Moreoever, the mechanism via which DNA is physically repaired is, itself, not a straightforward process. There exist multiple types of DNA repair \cite{ref:cha} such as nucleotide excision repair, mismatch repair, homologous recombination.
It is also possible that the proteins that search for, and are able to sense, DNA lesions are involved in the repair process as well as the search process. This would mean that, after having located a target site, rather than being released back into the system to search for another target, a protein may stay bound to the target site \cite{ref:zho}. Finally, whether searching for a binding site for gene transcription or searching for a DNA lesion in order to trigger the repair process, there are typically multiple searchers looking for multiple targets.

Thus, there are a plethora of future modifications that can be made to the basic model we study. 
Encouragingly, some studies of diffusion with resetting have begun to  explore  such details, for example,  including a  resetting distribution \cite{ref:eva2},
including dynamics in the absorption process \cite{WEM13}, including multiple searchers in the analysis \cite{Gelenbe10}, and  adding a delay to the reset process \cite{RUK14, Reuveni16, EM19, PKR19,  BS20}.

\begin{acknowledgements}
GRC acknowledges the Higgs Centre for Theoretical Physics where this work began as an MSc project.
\end{acknowledgements}

\noindent {\bf Author Contribution Statement:}
GRC designed and carried out research and drafted sections of the paper. MRE designed and carried out research and wrote the paper.

\appendix

\section{Calculation of $\widetilde{q}(x,s)$}
\label{app}

In this appendix we solve equation (\ref{lap})  self-consistently.
As mentioned in the main text there are number of approaches one can take;
here we choose to use the forward master equation.

For $x<0$ the solution of the homogeneous equation is
\begin{equation}
\widetilde{q}(x,s) = ae^{-\alpha{x}}+be^{\alpha{x}}
\end{equation}
and for $x>0$
\begin{equation}
\widetilde{q}(x,s) = ce^{-\alpha{x}}+de^{\alpha{x}}
\end{equation}
where 
\begin{equation}\label{alp}
    \alpha^2=\frac{(r+s)}{D}
\end{equation}
and $a$, $b$, $c$, $d$ are constants to be determined from the boundary conditions.
The boundary conditions $\widetilde{q}(x_l,s)=\widetilde{q}(x_r,s)=0$  give
\begin{eqnarray}
0 &=& ae^{-\alpha{x_l}}+be^{\alpha{x_l}}\\
0 &=&  ce^{-\alpha{x_r}}+de^{\alpha{x_r}}\,.
\end{eqnarray}
Continuity of $\widetilde{q}(x,s)$ at $x=0$ yields
\begin{equation}
    a+b=c+d\,,
\end{equation}
and the discontinuity in the derivative, obtained by integrating over the delta function,
\begin{equation}
    D\left[\frac{\partial\widetilde{q}(x,s)}{\partial{x}}\right]^{0^+}_{0^-}=-E\, ,
\end{equation}
implies
\begin{equation}
    -c+d+a-b=-\frac{E}{\alpha{D}}\; .
\end{equation}
There are now four simultaneous equations for  $a,b,c,d$ interms of $E$, the solution of which is
\begin{eqnarray}
a &=& \frac{E}{2\alpha{D}}e^{2\alpha{x_l}}\left(\frac{1-e^{2\alpha{x_r}}}{e^{2\alpha{x_r}}-e^{2\alpha{x_l}}}\right)\\[1ex]
b &=& \frac{-E}{2\alpha{D}}\left(\frac{1-e^{2\alpha{x_r}}}{e^{2\alpha{x_r}}-e^{2\alpha{x_l}}}\right)\\[1ex]
c &=& \frac{E}{2\alpha{D}}e^{2\alpha{x_r}}\left(\frac{1-e^{2\alpha{x_l}}}{e^{2\alpha{x_r}}-e^{2\alpha{x_l}}}\right)\\[1ex]
d &=& \frac{-E}{2\alpha{D}}\left(\frac{1-e^{2\alpha{x_l}}}{e^{2\alpha{x_r}}-e^{2\alpha{x_l}}}\right)
\end{eqnarray}
The self-consistent solution for $E$ is found  using  (\ref{Edef})
\begin{equation}
E = \frac{(r+s)(e^{\alpha{x_r}}+e^{\alpha{x_l}})}{r(e^{\alpha({x_r}+{x_l})}+1) + s(e^{\alpha x_r}+e^{\alpha {x_l}})}\;.
\end{equation}
\subsection*{Case $s=0$}
In the case $s=0$, which we will consider from now on, the expression for $E$ simplifies to
\begin{equation}
E = \frac{(e^{\alpha_0{x_r}}+e^{\alpha_0{x_l}})}{e^{\alpha_0({x_r}+{x_l})}+1}\;,
\end{equation}
where
\begin{equation}
\alpha_0 = \left( \frac{r}{D}\right)^{1/2}\;.
\end{equation}
Substituting $E$ in to $a,b,c,d$ then yields
\begin{eqnarray}
a &=& -\frac{e^{\alpha_0{x_l}}\sinh \alpha_0 x_r}{2\alpha_0{D}(\sinh{\alpha_0{x_r}}-\sinh{\alpha_0{x_l}})}\\
b &=& \frac{e^{-\alpha_0{x_l}}\sinh \alpha_0 x_r}{2\alpha_0{D}(\sinh{\alpha_0{x_r}}-\sinh{\alpha_0{x_l}})}\\
c &=&  -\frac{e^{\alpha_0{x_r}}\sinh \alpha_0 x_l }{2\alpha_0{D}(\sinh{\alpha_0{x_r}}-\sinh{\alpha_0{x_l}})}\\
d &=&  \frac{e^{-\alpha_0{x_r}} \sinh \alpha_0 x_l}{4\alpha_0{D}(\sinh{\alpha_0{x_r}}-\sinh{\alpha_0{x_l}})}\;.
\end{eqnarray}
Hence,  the expression  for $\widetilde{q}(x,0)$ reads
\begin{eqnarray}
    \widetilde{q}(x,0)&=&\frac{\sinh \alpha_0 x_r \sinh \alpha_0 (x-x_l)}{\alpha_0{D}(\sinh{\alpha_0{x_r}}-\sinh{\alpha_0{x_l}})} \;\;\mbox{for} \; x<0 \\
\widetilde{q}(x,0)&=& \frac{\sinh \alpha_0 x_l \sinh \alpha_0 (x-x_r)}{\alpha_0{D}(\sinh{\alpha_0{x_r}}-\sinh{\alpha_0{x_l}})}\;\; \mbox{for}\; x>0.
\end{eqnarray}

\end{document}